\documentclass[twocolumn,showpacs,preprintnumbers,amsmath,amssymb]{revtex4}
\pdfoutput=1
\usepackage{graphicx}% Include figure files
\usepackage{dcolumn}% Align table columns on decimal point
\usepackage{bm}% bold math
\usepackage{epsfig}
\usepackage{amsfonts,amsmath,amssymb,bm}

\begin{document}

\newcommand{\NB}{N\beta}
\newcommand{\EF}{E_F}

\title{Magnetic state electrical readout of Mn$_{12}$ molecules}

\author{C.D.~Pemmaraju, I.~Rungger and S.~Sanvito}
\affiliation{School of Physics and CRANN, Trinity College, Dublin 2, Ireland}

\begin{abstract}
We demonstrate that the different magnetic states of a Mn$_{12}$ molecule can be distinguished 
in a two-probe transport experiment from a complete knowledge of the current-voltage curve. 
Our results, obtained with state-of-the-art non-equilibrium transport methods combined with 
density functional theory, demonstrate that spin configuration-specific negative differential resistances 
(NDRs) appear in the $I$-$V$ curves. These originate from the interplay between electron localization
and the re-hybridization of the molecular levels in an external electric field and allow the detection
of the molecule's spin-state.
\end{abstract}

\date{\today}

\maketitle

The possibility of addressing and manipulating single spins represents the ultimate limit for the 
magnetic data storage industry. It may also lead to a technological platform for the integration of logic 
and data storage functionalities on the same device. A crucial aspect of any single spin-device is the 
development of a protocol for the electrical readout of the magnetic state of the device. 
Here we explore the possibility of inferring the spin configuration of a magnetic molecule by
a detailed knowledge of the $I$-$V$ characteristic of a two-probe device having the molecule as
resistive element. 

In particular we consider the case of Mn$_{12}$ sandwiched between non-magnetic electrodes 
and we compare the electrical response of two possible spin configurations of the molecule. 
First we look at the ground state (GS), where the spins of the eight peripheral Mn$^{3+}$ ions 
align antiparallel to those of the inner Mn$^{4+}$ to give an overall $S=10$ spin 
state \cite{Robertabook}. The second is a spin-flip (SF) configuration, in which the spin 
directions of one Mn$^{3+}$ and one Mn$^{4+}$ are reversed with respect to those in the
GS. The total spin-projection for such a state is $S=9$. The most notable difference between
the $I$-$V$ curves obtained for the different spin states is the much higher low-bias current 
of the GS configuration and the presence of NDRs, which are specific to the spin state.
These are a consequence of orbital re-hybridization under bias, which causes a highly non-linear 
bias-dependent coupling of the molecular levels to the electrodes. We predict that this is a
general feature of molecular junctions characterized by closely spaced orbital multiplets, such
as those appearing in magnetic molecules. Importantly, since both the orbital symmetry and
their localization depend on the molecule's spin-state, we expect different $I$-$V$ fingerprints 
for different magnetic configurations. This means that the overall molecular magnetic configuration 
is readable entirely from a single non-spin-polarized current readout.

Two experiments have recently measured the $I$-$V$ curve of three-terminal devices using
Mn$_{12}$ as the resistive element \cite{VdZ,Ralph}, in both cases revealing features associated to
the magnetic state of the molecule. NDRs have been ascribed to selection-rule forbidden 
transitions between different charging states \cite{Theo1,Theo2}, with the theory being
based on model Hamiltonians and parameters extracted from Mn$_{12}$ single crystal properties. 
A direct quantitative analysis of the results is complicated by the large variation in $I$-$V$s from 
device to device and by the low device production yield. This stems from the general fragility of Mn$_{12}$ 
in single molecule form on surfaces against fragmentation \cite{XAS} and redox 
\cite{Roberta1}. The spectroscopical detection of Mn$^{2+}$ in a Mn$_{12}$ monolayer  
\cite{XAS,Roberta1} is proof of such fragility. Furthermore, when the molecule remains
intact without changing its oxidation state, still the magnetic response is rather sensitive
to the surface environment and different from that of a single crystal. Most notably the anisotropy 
disappears \cite{Roberta2}, making the interpretation of the transport features in terms of 
selection rule-forbidden transitions more complicated.
Nevertheless, these pioneering transport measurements provide an indirect evidence for the 
different spin states of the molecule. Interestingly such evidence is obtained without using 
spin-polarized currents as in spin-polarized scanning tunnel microscopy (STM) \cite{STM1,STM2}
and so may represent an intriguing alternative for single-spin detection.

We begin our analysis by presenting the electronic structure of the archetypal Mn$_{12}$ acetate cluster 
[Mn$_{12}$O$_{12}$(CH$_3$COO)$_{16}$(H$_2$O)$_4$] both in the GS and in the SF configuration.
The calculations are performed with density functional theory (DFT) using the generalized
gradient approximation (GGA) \cite{PBE} as numerically implemented
in the pseudo-potential, local orbital basis set, {\sc siesta} code \cite{Siesta}. 
Details of the numerical method are presented as supplementary material.
The density of states (DOS) 
and the associated local DOS for a few of the molecular orbitals around the Fermi energy ($E_\mathrm{F}$) 
are presented in figure~\ref{Fig1}. Clearly the DOS is entirely dominated by the Mn$^{3+}$ $d$ shells  
(see also supplementary materials), with an energy gap between the highest 
occupied molecular orbital (HOMO) and the lowest unoccupied molecular orbital (LUMO) of about 0.3~eV. 
This is considerably smaller than the estimated experimental gap (1.4~eV \cite{Lic}) as expected 
from the self-interaction contributions \cite{ASIC} to the GGA functional. However such a reduced gap 
has little qualitative relevance for the low-bias $I$-$V$, since the nature of the HOMO manifold is
insensitive to the choice of the functional \cite{Lic}.

In the GS the DOS is entirely spin-polarized around $E_\mathrm{F}$ with a triply degenerate 
HOMO followed by a singly degenerate HOMO-1. Such a degeneracy is lifted in the SF configuration
where the HOMO is formed by four closely spaced Mn$^{3+}$ levels, one of which has the opposite spin.
Overall the total DOS (not considering the spin polarization) is rather similar in the two cases,
suggesting that an electrical measurement, which is not sensitive to the spin direction, will hardly be
able to distinguish between them. However there is an important difference between the two configurations.
This is the degree of localization of the various HOMO levels. In the GS the local DOS shows amplitude 
uniformly distributed around all the eight Mn$^{3+}$ ions, while in the SF state the minority HOMO is 
extremely localized around the flipped Mn$^{3+}$ ion. This has a profound influence over the electrical response
of the molecule.
\begin{figure}[ht]
\epsfxsize=7.5cm
\centerline{\epsffile{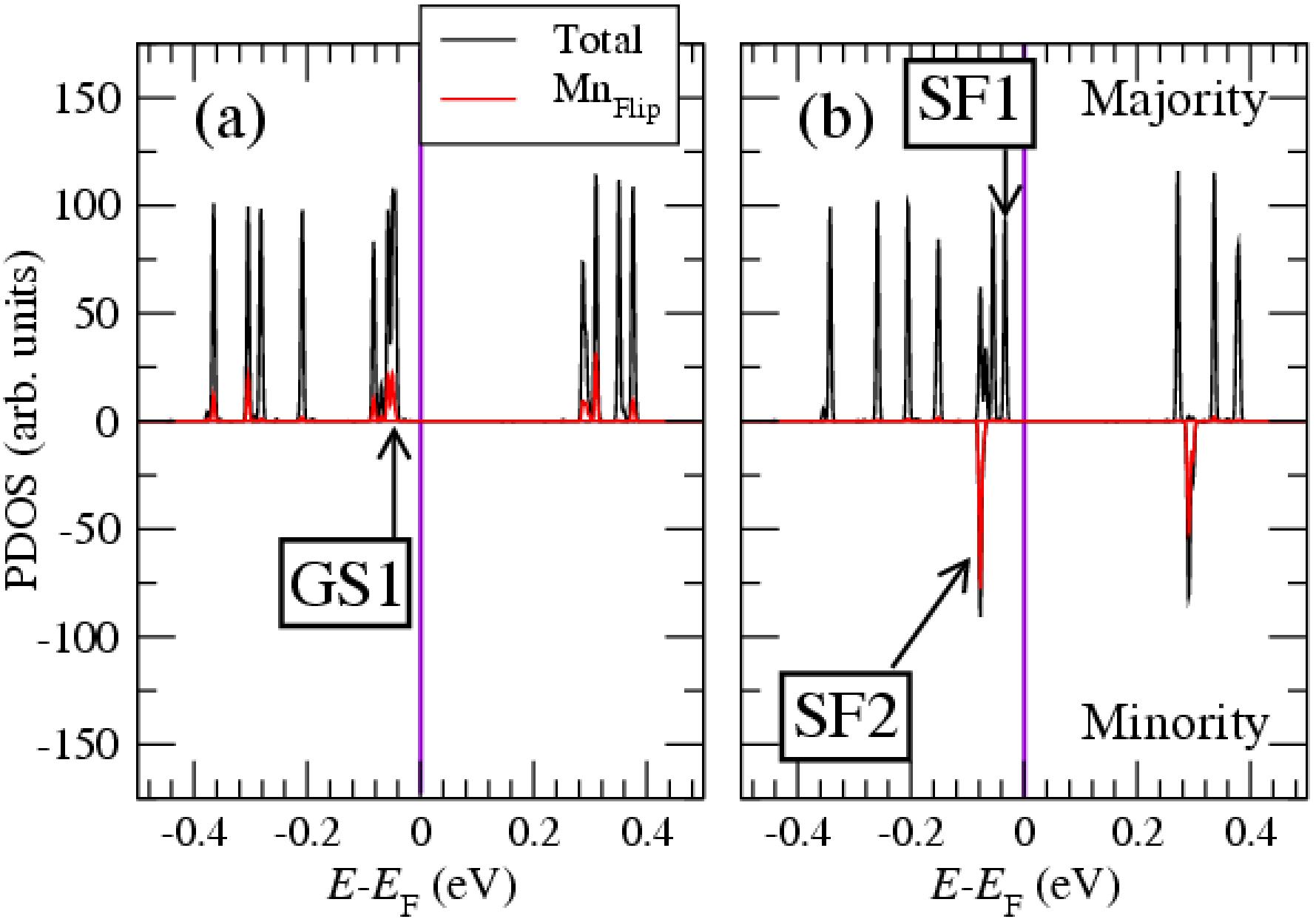}}
\centerline{
GS1{\epsfxsize=2.1cm
\epsffile{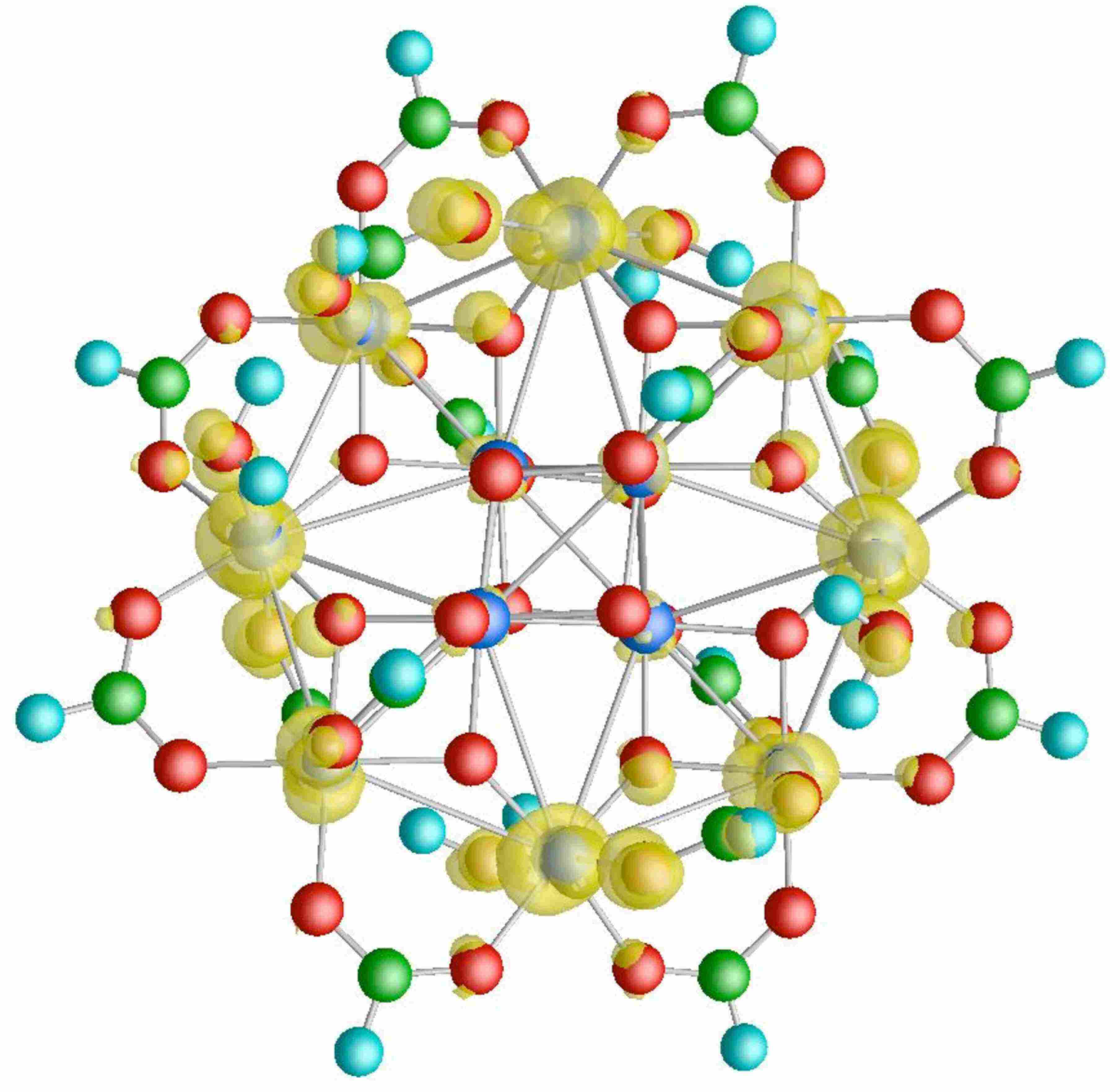}}
SF1{\epsfxsize=2.1cm
\epsffile{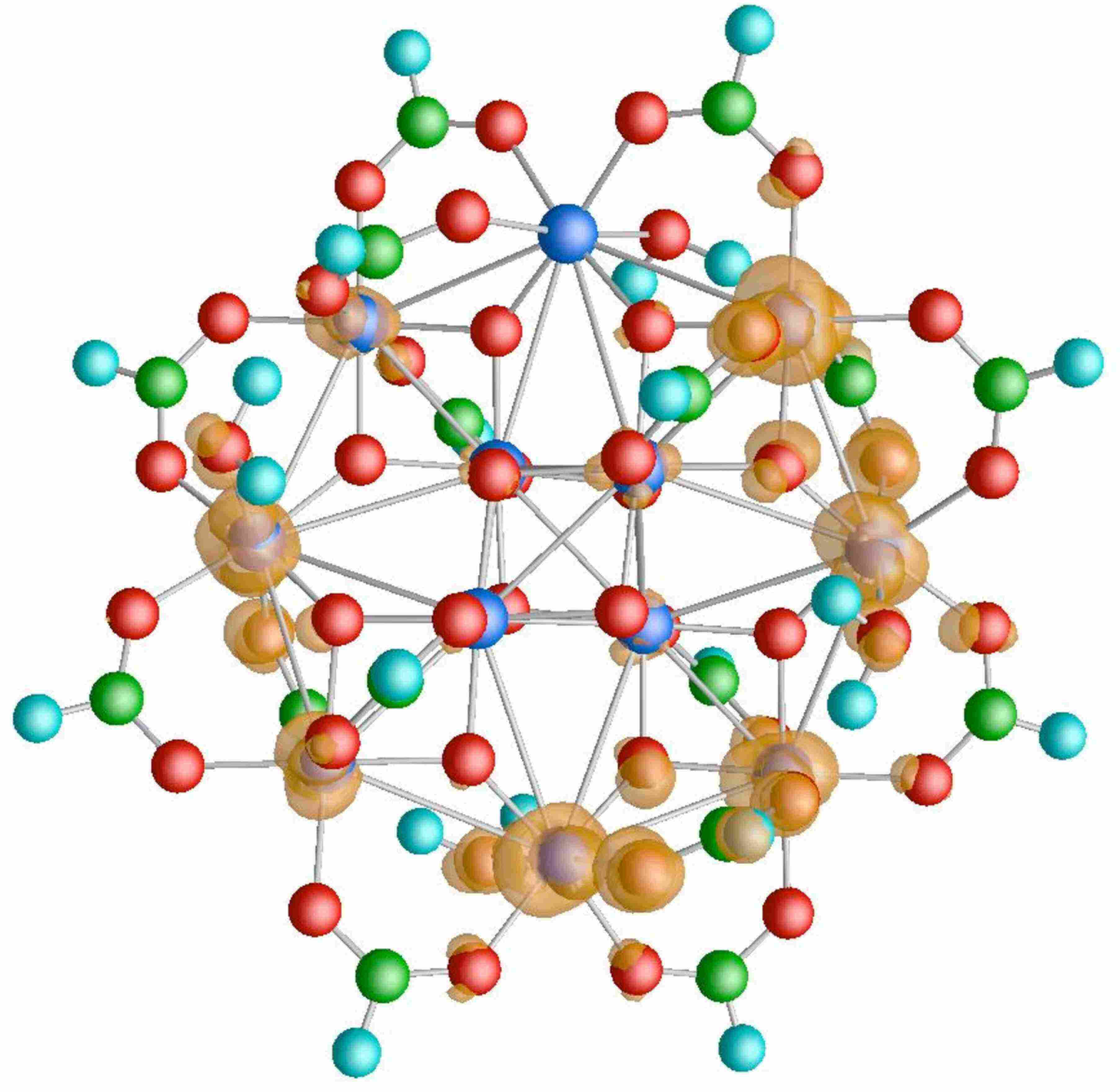}}
SF2{\epsfxsize=2.1cm
\epsffile{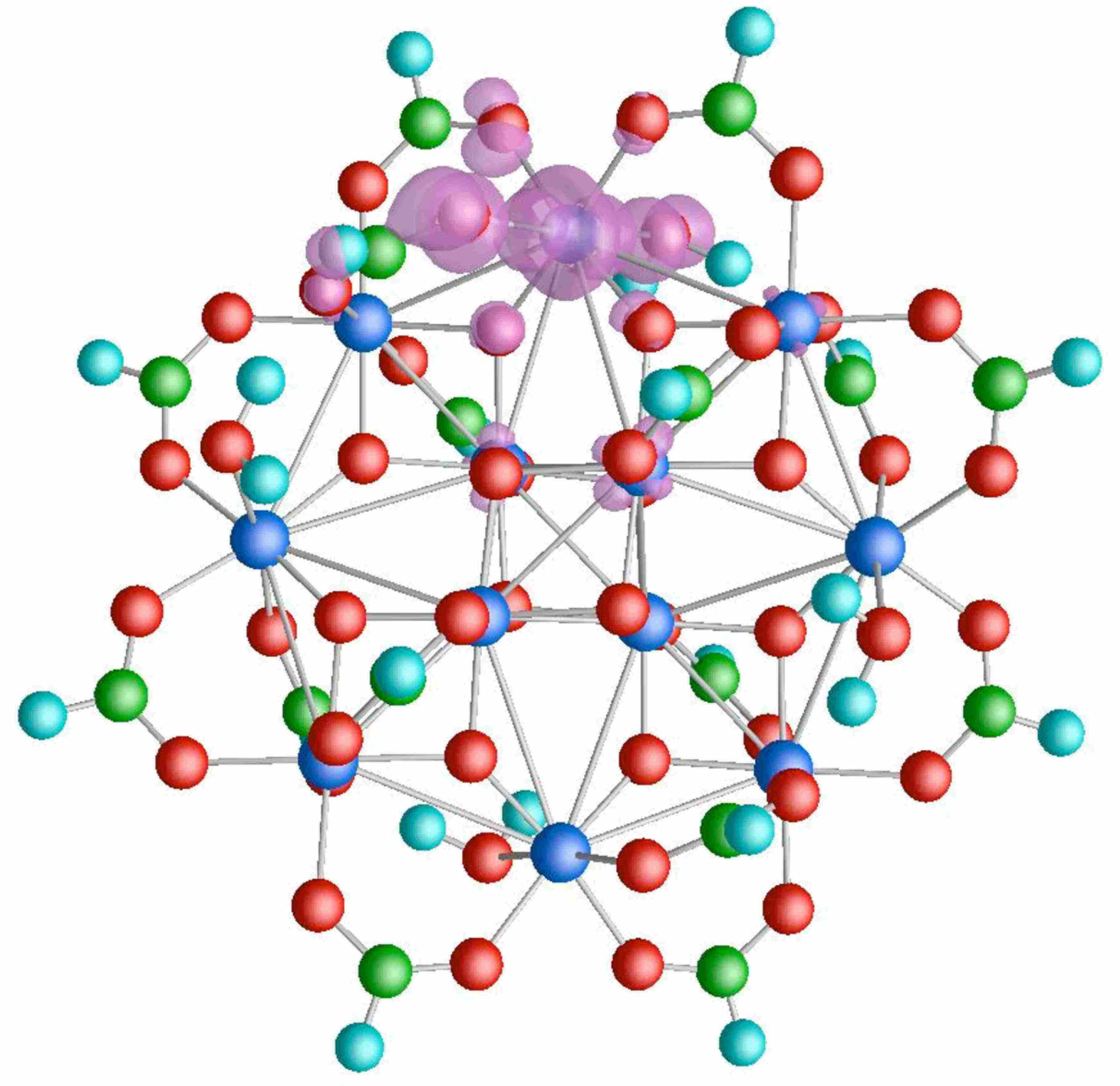}}}
\caption{Density of states around the Fermi level (purple vertical line) for 
[Mn$_{12}$O$_{12}$(CH$_3$COO)$_{16}$(H$_2$O)$_4$]  in the ground state (a) and the spin-flip state (b). 
Mn$_\mathrm{Flip}$ indicates the Mn$^{3+}$ ion whose spin direction was flipped when constructing the 
SF state. The lower panels display charge density isosurfaces (local density of states) respectively for the GS HOMO (GS1), and 
for both the HOMO (SF1) and the HOMO-4 (SF2) of the SF configuration. Note that SF2 has opposite spin direction 
with respect to the remaining HOMOs around $E_\mathrm{F}$. Color code: Blue=Mn, Red=O, Green=C, light blue=H.}
\label{Fig1}
\end{figure}
\\ 

We construct a Mn$_{12}$-based two terminal molecular device by sandwiching Mn$_{12}$ in between
two Au (111) surfaces. The molecule is anchored to Au by a thiol group attached to the C$_6$H$_4$
ligand, as in one of the published transport experiments \cite{VdZ}. In order to limit the lateral dimensions of the
unit cell for the transport calculations, we remove 12 of the possible 16 ligands and passivate the remaining 
bonds. The final simulation cell contains 620 atoms of which 480 are Au atoms of the electrodes, with the
basis set expanding over 2672 local orbitals for each spin direction. We have carefully checked that the first few HOMOs and LUMOs 
around $E_\mathrm{F}$ are not affected by the presence of the ligands or by the passivation. Since a full 
relaxation of the entire unit cell from DFT is numerically too demanding we first optimize the geometry of 
the molecule plus ligands and then we calculate the equilibrium bonding distance between the ligands 
and the hollow site of the Au (111) surfaces. The resulting cell is presented in the inset of figure \ref{Fig2}
(see also supplementary material).

Transport calculations are performed with the {\sc Smeagol} code \cite{Smeagol1,Smeagol2}, which combines
the non-equilibrium Green's function method (NEGF) \cite{datta,caroli}, with DFT as implemented in
{\sc Siesta} \cite{Siesta}. In this case the rather large size of the simulation cell requires a number
of technical solutions to boost the convergence. We have used our recently developed singularity removal 
procedure to evaluate accurately the leads self-energies \cite{Ivan}, and implemented a mesh 
refinement scheme for integrating efficiently the non-equilibrium charge density over energy (see supplementary material). 
The entire package was parallelized and optimized to run on a large scale computational facility.

The calculated $I$-$V$ curves, their spin decomposition and the associated differential conductances, $G(V)$,
are presented in figure~\ref{Fig2} for Mn$_{12}$ both in the GS and the SF configuration.
\begin{figure}[ht]
\epsfxsize=9.5cm
\centerline{\epsffile{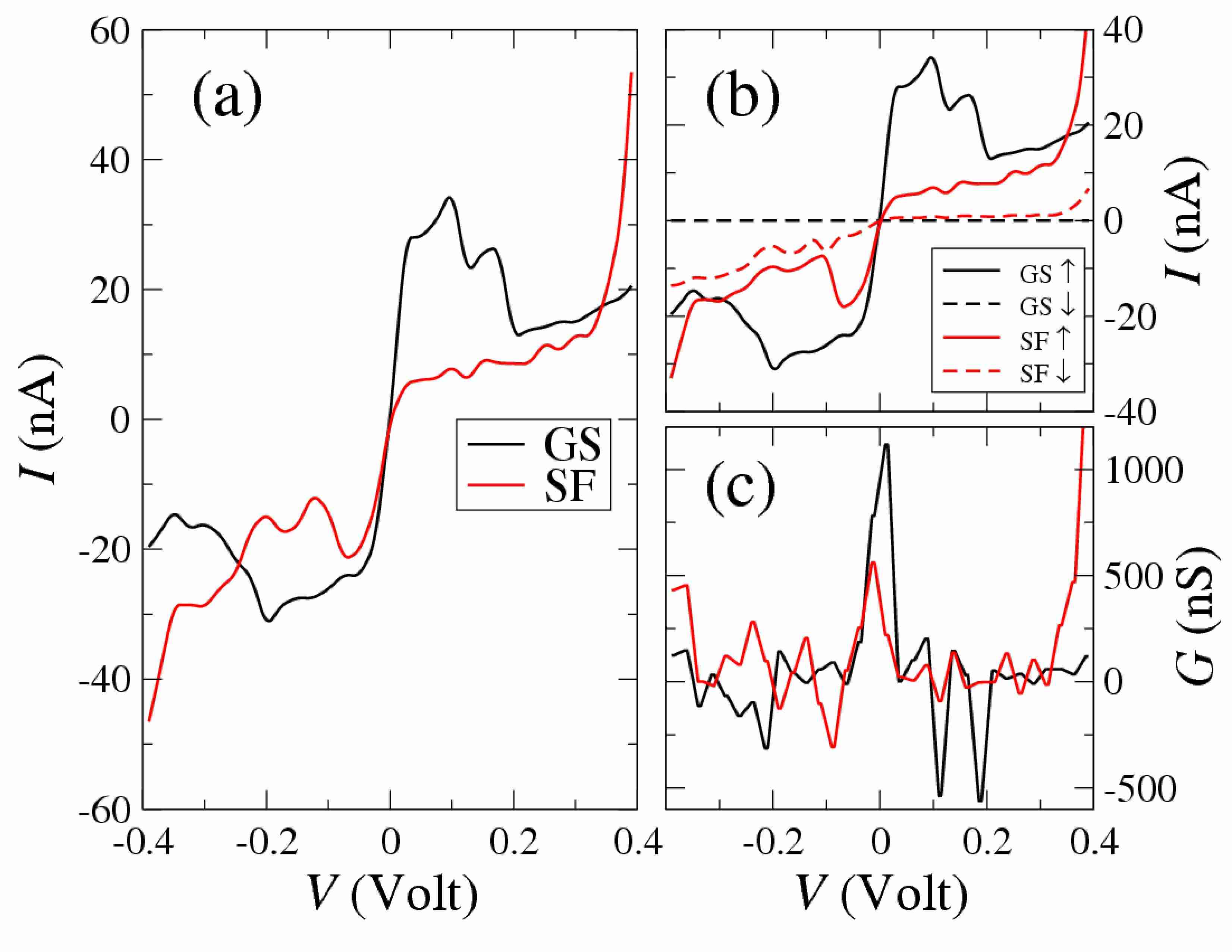}}
\vspace{-2.6cm}
\hspace{-2.6cm}
{\epsfxsize=2.9cm\epsffile{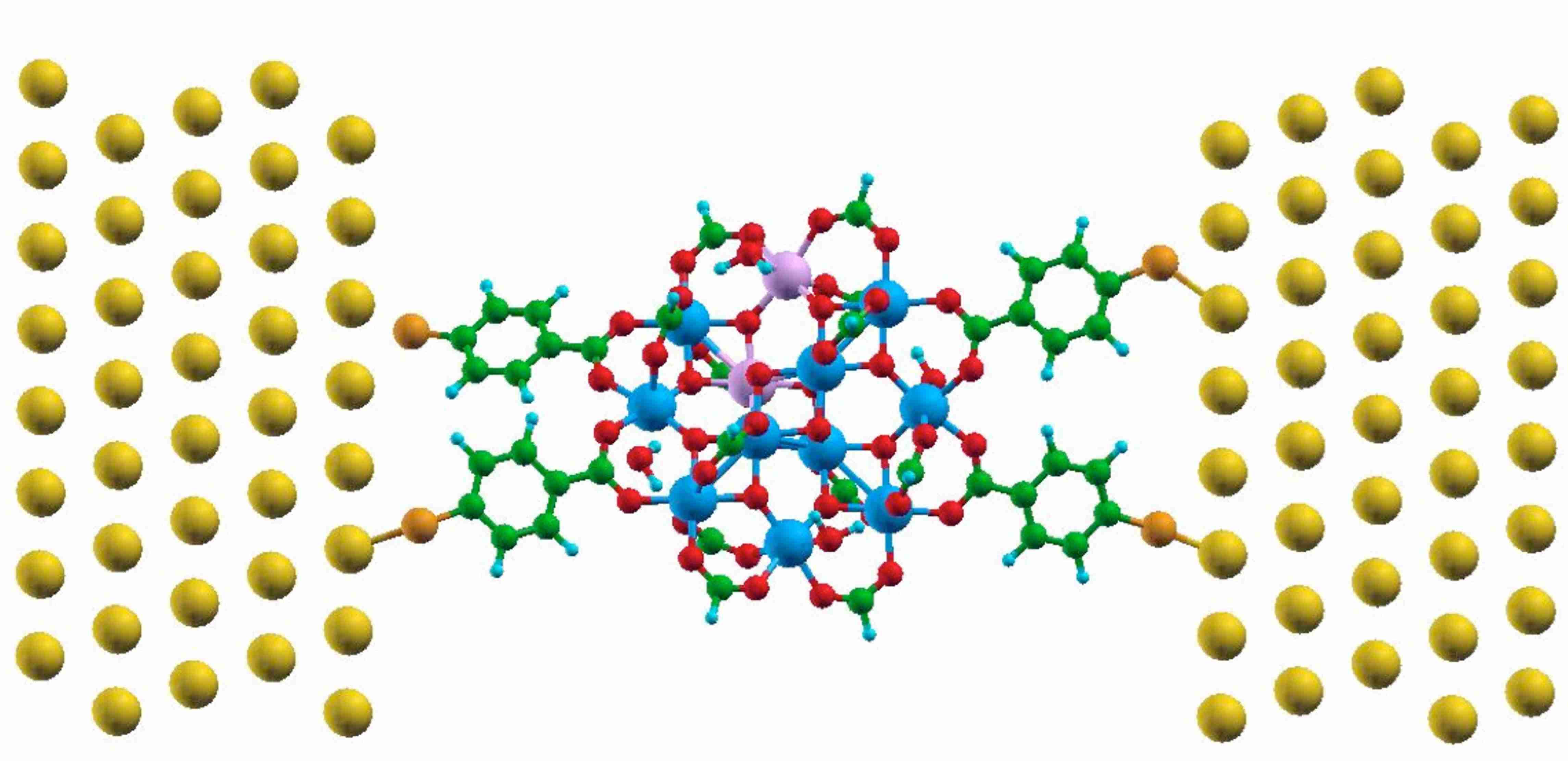}}
\vspace{0.6cm}
\caption{Transport properties of a Mn$_{12}$ two-probe device. The $I$-$V$ curves for both the GS and the SF
configuration are presented in (a), while (b) and (c) show the $I$-$V$ spin-decomposition and the differential
conductance respectively. The inset in panel (a) illustrates the simulation cell.}
\label{Fig2}
\end{figure}
The most notable feature is the rather stark difference between the GS and SF curves, which demonstrates
that a single current readout enables one to distinguish between the different spin states.
Importantly this does not require the readout of the two spin-components of the current, i.e. a spin-polarized
measurement, but it is simply obtained by comparing the general shape of the $I$-$V$ characteristics. 

In particular the distinctive fingerprint of the two magnetic configurations is in the presence of NDRs at 
different bias positions for the two configurations. These originate from the dynamics of the molecular 
energy levels under bias \cite{Alex}, their charging and their electrical polarization. The main concept
is that the different spin states of the molecule are associated to single particle levels with different orbital 
symmetry and hence with different response to an external electrical potential. Thus a particular spin configuration
translates into a distinct orbital configuration, which in turn is electrically readable. The details of how the various
molecular orbitals respond to the external potential are then determined by their coupling to the electrodes
and their charging energy.

These are investigated in figure~\ref{Fig3}, where we present the transmission coefficient as a function 
of energy, $T(E)$, calculated at different voltages for both the GS and the SF configuration. In order to understand our 
results let us recall the main features of electron transport through a generic molecular orbital $\psi_\alpha$. 
When $\psi_\alpha$ is weakly coupled to the electrodes the associated DOS, $D_\alpha(E)$, is 
approximated \cite{datta2} by
\begin{equation}
D_\alpha(E)=\frac{1}{2\pi}\frac{\gamma_\alpha}{(E-\varepsilon_\alpha)^2+(\gamma_\alpha/2)^2}\;,
\label{Lor}
\end{equation}
where $\gamma_\alpha=\gamma_\alpha^\mathrm{L}+\gamma_\alpha^\mathrm{R}$ and $\gamma_\alpha^\mathrm{L}/\hbar$
($\gamma_\alpha^\mathrm{R}/\hbar$) is the transmission rate for electron hopping from the left- (right-) hand side 
electrode to $\psi_\alpha$. The single particle energy $\varepsilon_\alpha$ depends on the charging state
of the molecule. In mean field approximation $\varepsilon_\alpha$ scales linearly \cite{Cormac} with the orbital 
occupation, $n_\alpha$, which in turn is determined by 
\begin{equation}
n_\alpha=\int_{-\infty}^{+\infty}\mathrm{d}E\:D_\alpha(E)
\frac{\gamma^\mathrm{L}_\alpha f^\mathrm{L}(E)+\gamma^\mathrm{R}_\alpha f^\mathrm{R}(E)}
{\gamma_\alpha}\:.
\label{Nsmb}
\end{equation}
Here $f^\mathrm{L}$ ($f^\mathrm{R}$) is the Fermi function evaluated at $E-\mu_\mathrm{L}$ ($E-\mu_\mathrm{R}$) 
with $\mu_\mathrm{L}$ ($\mu_\mathrm{R}$) the chemical potential of the left- (right-) hand side electrode 
($\mu_\mathrm{L}-\mu_\mathrm{R}=eV$). Finally the transmission coefficient is $T_\alpha(E)=2\pi\gamma_\alpha^\mathrm{L}
\gamma_\alpha^\mathrm{R}D_\alpha(E)/\gamma_\alpha$.

\begin{figure*}[h]
\centerline{\epsfxsize=9.5cm\epsffile{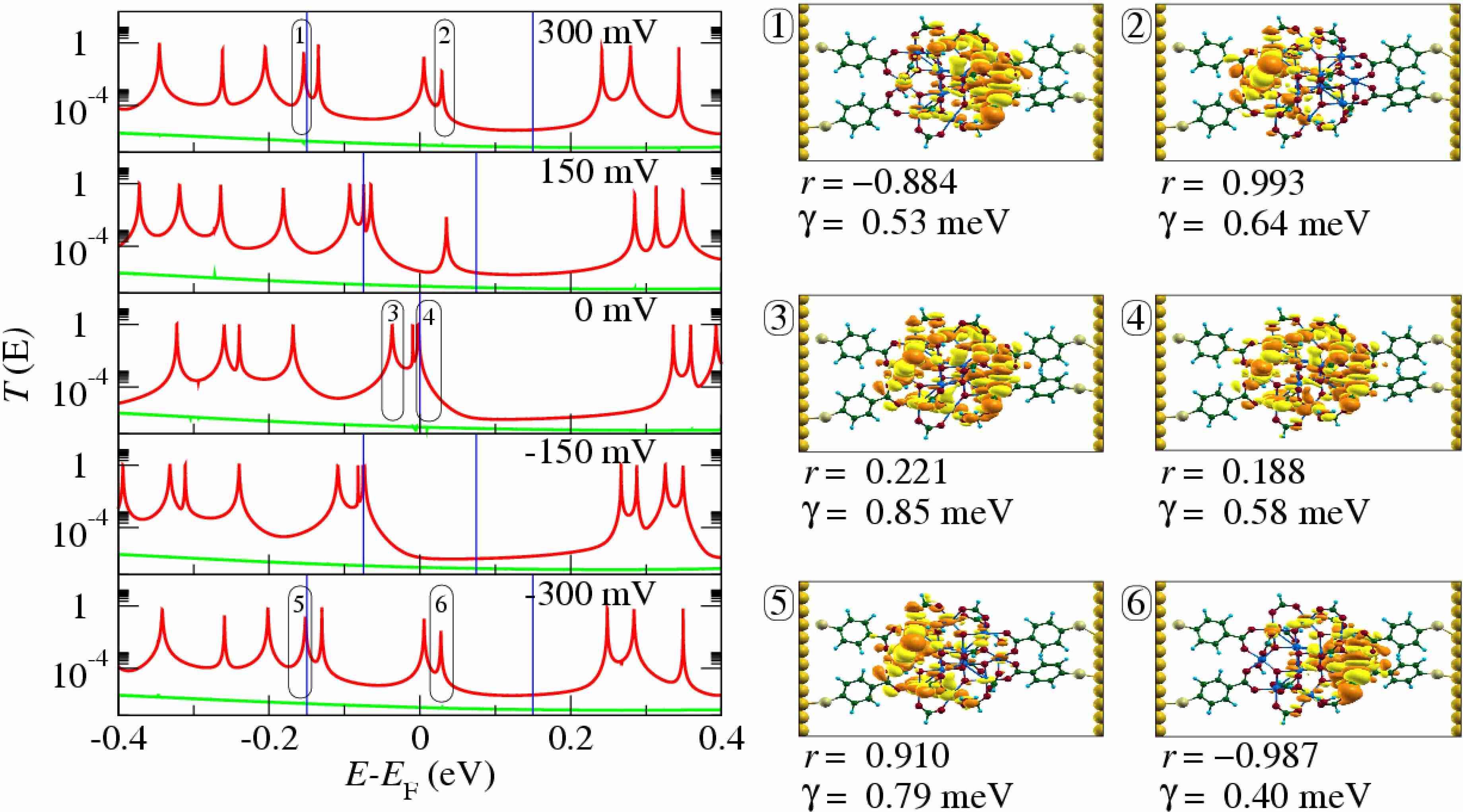}\hspace{0.5cm}\epsfxsize=9.5cm\epsffile{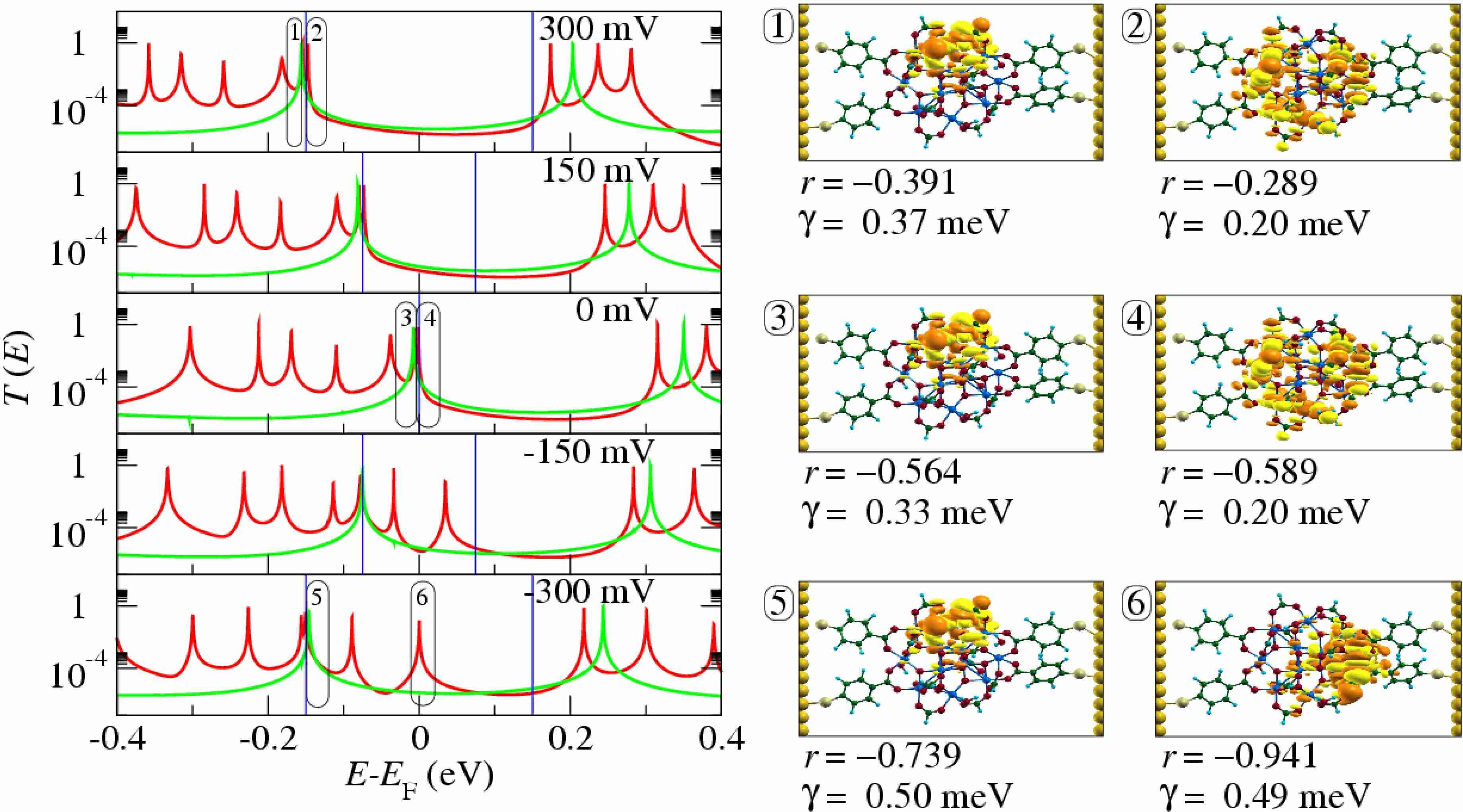}}
\caption{Transmission coefficient, $T(E)$ as a function of bias for both the GS (left) and the SF configuration (right).
The isosurfaces display the wave-functions of a number of molecular levels at different $V$. For these we also report
the value of both $r_\alpha$ and $\gamma_\alpha$. Note the dramatic re-hybridization with bias leading to highly 
asymmetric wave-functions.}
\label{Fig3}
\end{figure*}
The expression (\ref{Nsmb}) establishes that the steady state current is a 
balance of electron fluxes to and from both the electrodes. Thus if a molecular level is more strongly coupled
to one of the electrodes, as in an STM geometry, its single particle energy will be pinned to the chemical 
potential of that electrode, and the corresponding transmission peak $T(\varepsilon_\alpha)$ will
move in energy as a function of bias accordingly.
In contrast, when $\gamma_\alpha^\mathrm{L}\approx\gamma_\alpha^\mathrm{R}$, then $T_\alpha(\varepsilon_\alpha)$
as a function of $V$ is only determined by the level charging energy. Crucially in both the situations
the level broadening $\gamma_\alpha$ does not depend on bias and the width of the transmission
peak does not change with $V$. As a consequence $I(V)$ is solely determined by the {\it position} of the
transmission peaks. Since upon increasing the bias charging alone cannot expel from the bias window 
a molecular level which is already inside at a lower bias, no NDRs are expected. This is the most typical 
situation in molecular junctions.

In contrast the picture presented in figure~\ref{Fig3} is rather different. Let us take the GS as an example. 
At $V=0$ the first four HOMOs are less then 50~meV from $E_\mathrm{F}$ and approximately 
300~meV away from the first of the LUMO, hence they determine the low bias current. At positive bias
the current is initially determined by the first HOMO, which is pinned at $E_\mathrm{F}$. The HOMO moves 
into the bias window at about 100~mV and then roughly maintains its energy position at any larger 
positive $V$. Notably for $V<100$~mV the height of $T(\varepsilon_\mathrm{HOMO})$ is near to 1, 
suggesting that such a state is equally coupled to the electrodes. 
However when the level enters the bias window $T(\varepsilon_\mathrm{HOMO})$
reduces dramatically (note that the scale of Fig.~\ref{Fig3} is logarithmic), indicating that now the
coupling to the electrodes has changed. In general the magnitude of $T(\varepsilon_\alpha)$
reduces if $\gamma_\alpha^\mathrm{L}\ne\gamma_\alpha^\mathrm{R}$.
We then conclude that for the HOMO the coupling to the electrodes becomes asymmetric with bias. 

We can extract a quantitative parameter describing the dynamic coupling between the 
electrodes and the molecular orbitals as a function of bias from the spectral representation of the 
non-equlibrium Green's function. For each molecular level $\psi_\alpha$, one can 
evaluate the contributions to the imaginary part of the self-energy originating respectively from the
electronic coupling to the left ($\gamma_\alpha^\mathrm{L}$) and right ($\gamma_\alpha^\mathrm{L}$) 
electrode. In addition to the total broadening $\gamma_\alpha$ we now 
can define the coupling asymmetry parameter $r_\alpha=({\gamma_\alpha^\mathrm{L}-\gamma_\alpha^\mathrm{R}})/
({\gamma_\alpha^\mathrm{L}+\gamma_\alpha^\mathrm{R}})$. This is +1 (-1) for molecular orbitals more
strongly coupled to the left (right) contact and 0 for coupling equally strong to both
the electrodes. 

\begin{figure*}[h]
\centerline{\epsfxsize=16.5cm\epsffile{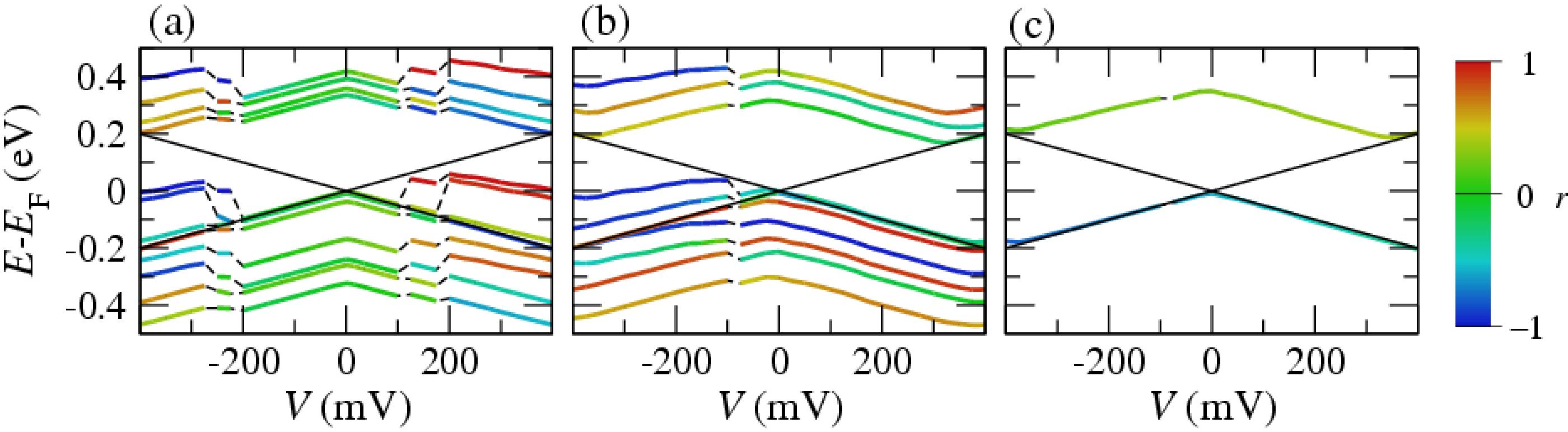}}
\caption{Trace of $r_\alpha$ as a function of energy and bias. The colored lines denote the position of
the peaks in the transmission coefficient and the color encodes the magnitude of $r_\alpha$. The two solid
straight black lines mark the boundaries of the bias window. Panel (a) is for the GS and majority spins,
panel (b) is for the SF configuration and majority spins, panel (c) is for the SF configuration and minority spins.
Color code: red $r_\alpha=1$, green $r_\alpha=0$, blue $r_\alpha=-1$.}
\label{Fig4}
\end{figure*}

\begin{figure*}[h]
\centerline{\epsfxsize=16.5cm\epsffile{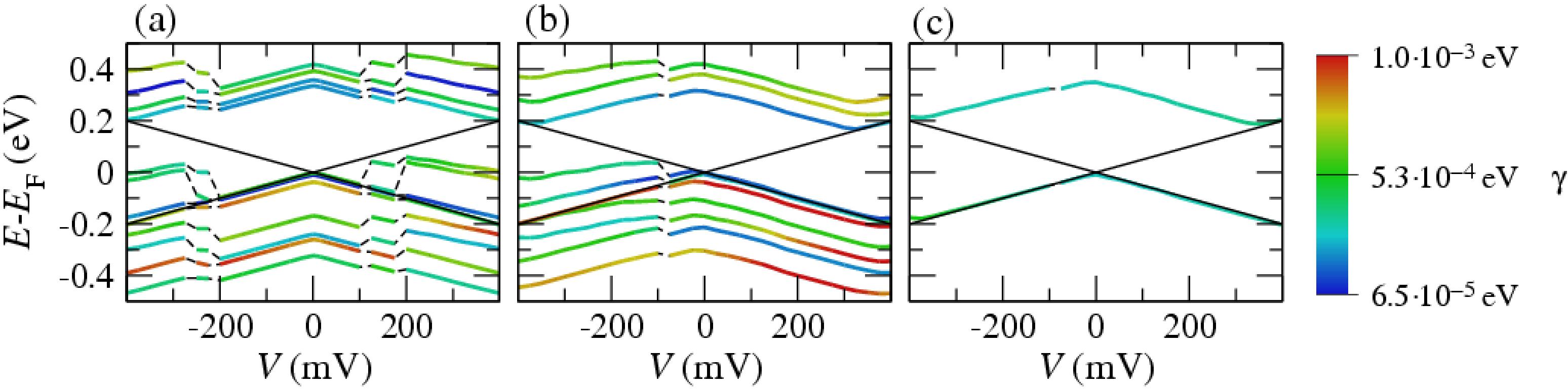}}
\caption{Trace of $\gamma_\alpha$ as a function of energy and bias. The colored lines denote the position of
the peaks in the transmission coefficient and the color encodes the magnitude of $\gamma_\alpha$. The two solid
straight black lines mark the boundaries of the bias window. Panel (a) is for the GS and majority spins,
panel (b) is for the SF configuration and majority spins, panel (c) is for the SF configuration and minority spins.
Color code: red $\gamma_\alpha=10^{-3}$~eV, blue $\gamma_\alpha=6.5\cdot10^{-5}$~eV.}
\label{Fig5}
\end{figure*}

In figures~\ref{Fig4} and \ref{Fig5} we trace $\varepsilon_\alpha$ from the position of the various peaks in the 
transmission coefficients as a function of $V$. The colors encode the value of $r_\alpha$ (in Fig.~\ref{Fig4})
and $\gamma_\alpha$ (in Fig.~\ref{Fig5}), and the two straight black lines mark the bias window. Let us
start again from the GS [panels (a)]. As mentioned before for small $V$ the first HOMO is pinned 
to the lower bound of the bias window and, confirming our initial guess, it is roughly equally coupled to the
electrodes [Fig.~\ref{Fig4}(a)]. However for $V\sim100$~mV it suddenly jumps into
the bias window. In doing so a remarkable effect takes place: its coupling to the leads 
becomes extremely asymmetric, $r_\alpha\sim1$, while $\gamma_\alpha$ remains approximately constant.
Considering the fact that the contribution to the current from a transmission peak, $I_\alpha$, is approximately 
\begin{equation}
I_\alpha=2\pi\frac{e}{h}\frac{\gamma_\alpha(1-r_\alpha^2)}{4}\;,
\end{equation}
we conclude that as soon as the peak enters the bias window the total current actually decreases.
This creates the NDR at 100~mV shown in figure~\ref{Fig2}(c). Similarly the second HOMO enters the
bias window at 200~mV, again with a drastic change of $r_\alpha$. This produces the second
NDR at 200~mV.

For $V<0$ the situation is analogous although quantitatively different. Again at low bias the HOMO follows 
the lower boundary of the bias window with an almost equal coupling to the two electrodes. Then at about $-200$~mV it suddenly
enters the bias window together with the HOMO-2. Similarly to the $V>0$ case entering into the
bias window is accompanied by a drastic change in coupling. This time however $r_\alpha\sim-1$, 
indicating that the molecular level is now more strongly coupled to the right-hand side
electrode. As a result the NDRs at negative $V$ appears only at $-200$~mV.

We now turn our attention to the SF configuration. The most notable feature is represented by the 
minority HOMO, for which the coupling to the leads is almost insensitive to the bias 
and it is rather equally strong to both the electrodes [see Fig.\ref{Fig4}(c) and Fig~\ref{Fig5}(c)]. Therefore 
the level always follows the lower boundary of the bias window and remains occupied at any voltage,
thus resulting in an almost linear contribution to the $I$-$V$. The majority component of the current instead
displays a behaviour qualitatively similar to that of the GS. This time however the first two HOMOs enter
the bias window together but only at negative voltages (around $-100$~mV), again in correspondence to an 
increased asymmetry of the electronic coupling with the electrodes [$r_\alpha\sim-1$, see Fig.~\ref{Fig4}(b)].
This generates the NDR at $-100$~mV.

Finally we answer the following question: what is the mechanism behind the bias-dependence of $r_\alpha$? 
Let us consider the isosurfaces of Fig.~\ref{Fig3}, displaying the electronic wave function of a given molecular level 
in presence of $V$. For example let us look at the panels 2, 4 and 6 for the GS. These represent the wave-function 
of the first HOMO respectively at 300~mV, 0 and $-300$~mV. Clearly as bias is applied, the wave function experiences a substantial
polarization, with the electron clouds changing from a uniform distribution at $V=0$ to one which is considerably
localized either to the left- or to the right-hand side of the junction. The molecular orbital distortion under bias
originates from molecular orbital re-hybridization and always involves a number of molecular states. For instance,
when the HOMO and HOMO-1 enter the bias window at $V>0$ the re-hybirdization involves the first several HOMOs,
as one can deduce from the fact that for most of them there is a change in $r_\alpha$ [see Fig.~\ref{Fig4}(a)]. 
Importantly for the SF configuration there is only one minority spin level around $E_\mathrm{F}$, this cannot hybridize
with any other state (unless a spin-flip event occurs), and therefore its coupling to the electrodes changes
little with bias. This fundamental aspect is what creates profoundly different $I$-$V$ curves in the two cases.

Crucially orbital re-hybridization results in the molecular levels entering the bias window always to display largely
asymmetric coupling to the electrodes. This allows them to enter the bias window and to conduct without charging the molecule, 
which remains neutral at any bias investigated. We therefore conclude that the observed NDRs are features in the $I$-$V$ 
occurring at energies lower than the typical charging energy of the molecule. Note that in general orbital re-hybridization may
not only lead to NDRs, but also to a sharp increase of $G(V)$. Importantly these are always low-energy satellite features 
to the main Coulomb blockade signal.

In conclusion we have demonstrated that the $I$-$V$ characteristics of Mn$_{12}$-based devices are sensitive to the 
internal magnetic state of the molecule, so that an electrical readout can distinguish between two different magnetic
configurations without the need of resolving the two spin components of the current. 
Such a sensitivity originates from the bias dependence of the electronic coupling between the molecular levels and 
the electrodes. In the case examined here this provides a novel mechanism for NDRs, which indeed are the
final fingerprints of a specific magnetic state of the molecule. Significantly this represents also a new mechanism for 
NDR, requiring a number of closely spaced molecular levels close to the electrodes' Fermi level. 
Single molecules magnet, with their $d$-electron manifold, appear as the ideal material system for such an
effect.

\small{

}

%\bibliography{Paper_1}% Produces the bibliography via BibTeX.

\begin{thebibliography}{99}

\bibitem{Robertabook} D.~Gatteschi, R.~Sessoli and J.~Villain, {\em Molecular Nanomagnets}, Oxford University Press (Oxford, 2006).

\bibitem{VdZ} H.B.~Heersche et al., 
%Electron Transport through Single Mn$_{12}$ Molecular Magnets,
Phys. Rev. Lett. {\bf 96}, 206801 (2006).

\bibitem{Ralph} M.-H.~Jo et al., 
%Signatures of molecular magnetism in single-molecule transport spectroscopy, 
Nano Lett. {\bf 6}, 2014 (2006).

\bibitem{Theo1}C.~Romeike, M.R.~Wegewijs and H.~Schoeller, 
%Spin quantum tunneling in single molecular magnets: fingerprints in transport spectroscopy of current and noise, 
Phys. Rev. Lett. {\bf 96}, 196805 (2006).

\bibitem{Theo2}C.~Romeike, M.R.~Wegewijs, M.~Ruben, W.~Wenzel and H.~Schoeller,
%Charge-switchable molecular magnet and spin blockade of tunneling, 
Phys. Rev. B {\bf 75}, 064404 (2007).

\bibitem{XAS}S.~Voss, M.~Fonin, U.~R\"udiger, M.~Burgert, U.~Groth, Yu.S.~Dedkov, 
%Electronic structure of Mn$_{12}$ derivatives on the clean and functionalized Au surface,
Phys. Rev. B {\bf 75}, 045102 (2007).

\bibitem{Roberta1}M.~Mannini, P.~Sainctavit, R.~Sessoli, C.~Cartier~dit~Moulin, F.~Pineider, M.-A.~Arrio, A.~Cornia, 
and D.~Gatteschi, 
%XAS and XMCD Investigation of Mn$_{12}$ Monolayers on Gold,
Chem. Eur. J. {\bf 14}, 7530 (2008).

\bibitem{Roberta2}M.~Mannini, F.~Pineider, P.~Sainctavit, L.~Joly, A.~Fraile-Rodr�guez, M.-A.~Arrio, 
C.~Cartier~dit~Moulin, W.~Wernsdorfer, A.~Cornia, D.~Gatteschi and R.~Sessoli, 
%X-Ray Magnetic Circular Dichroism Picks out Single-Molecule Magnets Suitable for Nanodevices,
Adv. Mater. {\bf 21}, 167 (2008).

\bibitem{STM1}F.~Meier, L.~Zhou, J.~Wiebe and R.~Wiesendanger, 
%Revealing Magnetic Interactions from Single-Atom Magnetization Curves,
Science {\bf 320}, 82 (2008).

\bibitem{STM2}C.F.~Hirjibehedin, C.-Y.~Lin, A.F.~Otte, M.~Ternes, C.P.~Lutz, B.A.~Jones and A.J.~Heinrich, 
%Embedded in a Surface Molecular Network Large Magnetic Anisotropy of a Single Atomic Spin,
Science {\bf 317}, 1199 (2007).

\bibitem{PBE}J.P.~Perdew, K.~Burke and M.~Ernzerhof, 
%Generalized Gradient Approximation Made Simple,
Phys. Rev. Lett. {\bf 77}, 3865 (1996).

\bibitem{Siesta}J.~M.~Soler, E.~Artacho, J.~D.~Gale, A.~Garcia, J.~Junquera, P.~Ordej\'on and D.~Sanchez-Portal, 
%The {\sc siesta} method for ab initio order-N materials simulation,
J. Phys. Condens. Matter {\bf 14}, 2745-2779 (2002).

\bibitem{Lic}D.W.~Boukhvalov, M.~Al-Saqer, E.Z.~Kurmaev, A.~Moewes, V.R.~Galakhov, L.D.~Finkelstein, 
S.~Chiuzb\"{a}ian, M.~Neumann, V.V.~Dobrovitski, M.I.~Katsnelson, A.I.~Lichtenstein, B.N.~Harmon, 
K.~Endo, J.~M. North and N.S.~Dalal,
%Electronic structure of a Mn$_{12}$ molecular magnet: Theory and experiment,
Phys. Rev. B {\bf 75} 014419 (2007).

\bibitem{ASIC}C.~Das~Pemmaraju, T.~Archer, D.~S\'anchez-Portal and S. Sanvito, 
%Atomic-orbital-based approximate self-interaction correction scheme for molecules and solids, 
Phys. Rev. B {\bf 75} 045101, (2007). 

\bibitem{Smeagol1}A.R. Rocha, V.M.~Garcia Suarez, S.W. Bailey, C.J. Lambert, J.~Ferrer and S.~Sanvito, 
%Spin and Molecular Electronics in Atomically-Generated Orbital Landscapes, 
Phys. Rev. B {\bf 73}, 085414, (2006).

\bibitem{Smeagol2}A.R. Rocha, V.M.~Garcia Suarez, S.W. Bailey, C.J. Lambert, J.~Ferrer and S.~Sanvito,
%Towards molecular spintronics,
Nature Materials, {\bf 4} 335, (2005).

\bibitem{datta}S. Datta, {\it Electronic Transport in Mesoscopic Systems}, (Cambridge University Press,
Cambridge, UK, 1995).

\bibitem{caroli}C.~Caroli, R.~Combescot, P.~Nozieres, and D.~Saint-Janes,
%A direct calculation of the tunneling current: IV. Electron-phonon interaction effects,
J. Phys. C {\bf 5}, 21 (1972).

\bibitem{Ivan}I.~Rungger and S.~Sanvito,
%Algorithm for the construction of self-energies for electronic transport calculations based on singularity elimination 
%and singular value decomposition,
Phys. Rev. B {\bf 78}, 035407 (2008).

\bibitem{Alex}A.~Reily Rocha and S.~Sanvito,
%Asymmetric I-V characteristics and magnetoresistance in magnetic point contacts,
Phys. Rev. B {\bf 70}, 094406 (2004).

\bibitem{datta2}S.~Datta,
%Electrical resistance: an atomistic view,
Nanotechnology {\bf 15}, S433 (2004).

\bibitem{Cormac}C.~Toher, A.~Filippetti, S.~Sanvito and K.~Burke, 
%Self-Interaction errors in density-functional calculations of electronic transport,
Phys. Rev. Lett. {\bf 95}, 146402 (2005).

\end{thebibliography}

\begin{thebibliography}{99}

\bibitem{Smeagol1}A.R. Rocha, V.M.~Garcia Suarez, S.W. Bailey, C.J. Lambert, J.~Ferrer and S.~Sanvito, 
%Spin and Molecular Electronics in Atomically-Generated Orbital Landscapes, 
Phys. Rev. B {\bf 73}, 085414, (2006).

\bibitem{Smeagol2}A.R. Rocha, V.M.~Garcia Suarez, S.W. Bailey, C.J. Lambert, J.~Ferrer and S.~Sanvito,
%Towards molecular spintronics,
Nature Materials, {\bf 4} 335, (2005).

\bibitem{Siesta}J.~M.~Soler, E.~Artacho, J.~D.~Gale, A.~Garcia, J.~Junquera, P.~Ordej\'on and D.~Sanchez-Portal, 
%The {\sc siesta} method for ab initio order-N materials simulation,
J. Phys. Condens. Matter {\bf 14}, 2745-2779 (2002).

\bibitem{Cormac}For a discussion about the possible bonding sites of the thiol group on Au (111) see for instance,
C.~Toher and S.~Sanvito, Phys. Rev. B {\bf 77}, 155402 (2008).

\end{thebibliography}

\section*{Acknowledgment}

We thank Greg Szluncheski for interesting discussions and Roberta Sessoli for having driven 
our attention towards Mn$_{12}$ and for having provided us with crystallographic data on
Mn$_{12}$.
This work is sponsored by Science Foundation of Ireland (07/IN.1/I945 and 07/RFP/PHYF235) 
and by the European Union (SpiDME EC-FP6 NEST). Computational resources have been 
provided by the HEA IITAC project managed by the Trinity Centre for High Performance Computing 
and by the Irish Center for High-End Computing.

\clearpage
%\newpage

%\pagestyle{fancy} \lhead{}
%\rhead{\small \sc Supplementary Materials \hspace{8.0cm} -- C.D. Pemmaraju et al.}
\begin{center}
\section*{Supplementary Materials} 
\end{center}

\section{Computational Method}

\begin{figure}[ht]
\centerline{\epsfxsize=7.9cm\epsffile{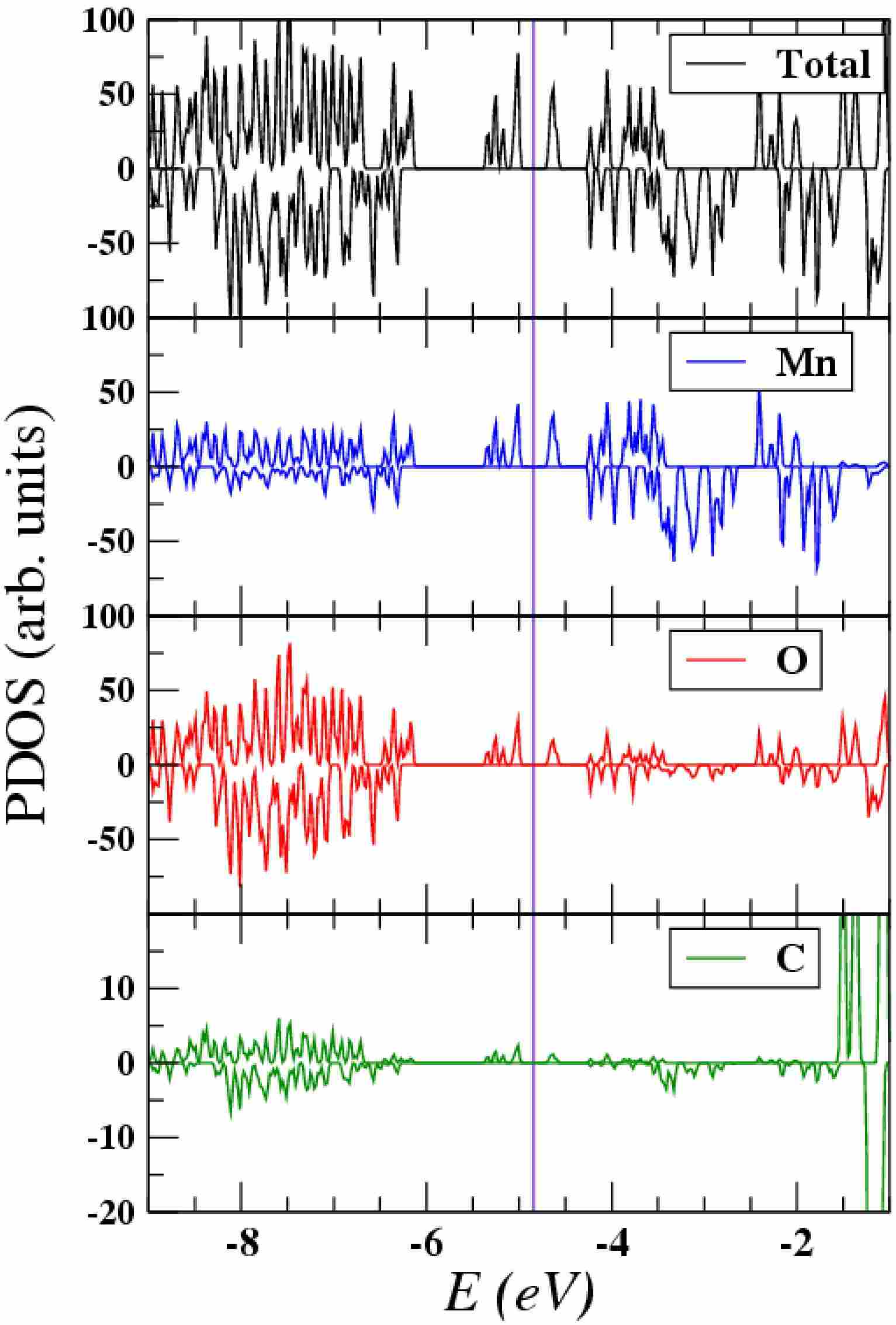}}
\caption{Orbital resolved density of states around the Fermi level (purple vertical line) for 
[Mn$_{12}$O$_{12}$(CH$_3$COO)$_{16}$(H$_2$O)$_4$]  in the $S=10$ ground state.
Note that the HOMO-LUMO gap is an intra-$d$ gap with states originating mainly from
Mn ions (mainly 3+, see main text). Note also the strong Mn-O hybridization both in the 
HOMO and LUMO.}
\label{Fig1}
\end{figure}

Calculations have been all performed with the DFT and NEGF {\sc smeagol} code \cite{Smeagol1,Smeagol2}.
We use standard scalar relativistic pseudopotentials with the following reference configurations: H 1$s^1$, 
C 2$s^2$2$p^2$, S 3$s^2$3$p^4$, O 2$s^2$2$p^4$, Mn 4$s^2$3$d^5$ and Au 6$s^1$. The atomic 
basis set is constructed as follows C: DZ-$s$, DZ-$p$; H: DZP-$s$; O: DZ-$s$, DZP-$p$, SZ-$d$; S: DZ-$s$, DZP-$p$, SZ-$d$;
Mn: DZP-$s$, SZP-$p$, DZ-$d$; Au: DZ-$s$ (the notation is SZ=single zeta, DZ=double zeta, P=polarized)\cite{Siesta}.
The real space grid has an equivalent cutoff energy of 400~Ry.
The charge density is obtained by splitting the integral of the Green's function (GF) into a contribution calculated over the
complex energy plane and one along the real axis \cite{Smeagol1}. The complex integral is performed over a uniform
mesh of 512 imaginary energies, while for the real part we have implemented a new mesh refinement algorithm,
necessary to integrate the extremely sharp features of the DOS. Such an algorithm consists in evaluating the integral 
over an initial coarse energy mesh after the GF has been artificially broadened. Then peaks in the GF are detected and
a denser energy mesh is generated around them. The next approximation is then calculated over this new grid after
the artificial broadening has been decreased. Such a procedure is repeated until no artificial broadening is left. 
Typically the final mesh includes 1000 energy points with the denser energy spacing being around $10^{-5}$~eV.

The hopping rates of a molecular orbital to the electrodes, $\gamma_\alpha^\mathrm{L}$ and $\gamma_\alpha^\mathrm{R}$, 
used in the analysis of the transmission coefficients, are calculated by using the spectral representation of the GF associated to the
scattering region
\begin{equation}
G=\sum_\alpha\frac{1}{E-(\varepsilon_\alpha-i{\gamma_\alpha}/{2})+i0^+}\psi_\alpha\tilde{\psi}_\alpha^\dagger\:,
\end{equation}
where $\psi_\alpha$ and $\tilde{\psi}_\alpha$ are the right and left eigenvectors of $H_\mathrm{eff}$ and 
$(\varepsilon_\alpha-i{\gamma_\alpha}/{2}$) are the associated eigenvalues. Finally 
\begin{equation}
H_\mathrm{eff}=H_\mathrm{M}+\Sigma_\mathrm{L}+\Sigma_\mathrm{R}\:,
\end{equation}
with $H_\mathrm{M}$ the Hamiltonian of the scattering region (molecules plus part of the leads) and $\Sigma_\mathrm{R}$
($\Sigma_\mathrm{L}$) is the self-energy for the right- (left-) hand side electrode \cite{Smeagol1}. The coupling constants
$\gamma_\alpha^n$ ($n$=L, R) are obtained as
\begin{equation}
\gamma_\alpha^n=\tilde{\psi}_\alpha^\dagger[i(\Sigma_n-\Sigma_n^\dagger)]\tilde{\psi}_\alpha\:.
\end{equation} 
One can easily demonstrate that with this definition $\gamma_\alpha=\gamma_\alpha^\mathrm{L}+\gamma_\alpha^\mathrm{R}$ 
and the orbital occupation of $\psi_\alpha$ is [equation (2) of the main text].
\begin{equation}
n_\alpha=\int_{-\infty}^{+\infty}\mathrm{d}E\:D_\alpha(E)
\frac{\gamma^\mathrm{L}_\alpha f^\mathrm{L}(E)+\gamma^\mathrm{R}_\alpha f^\mathrm{R}(E)}
{\gamma_\alpha}\:.
\label{Nsmb}
\end{equation}

\begin{figure}[ht]
\centerline{\epsfxsize=7.9cm\epsffile{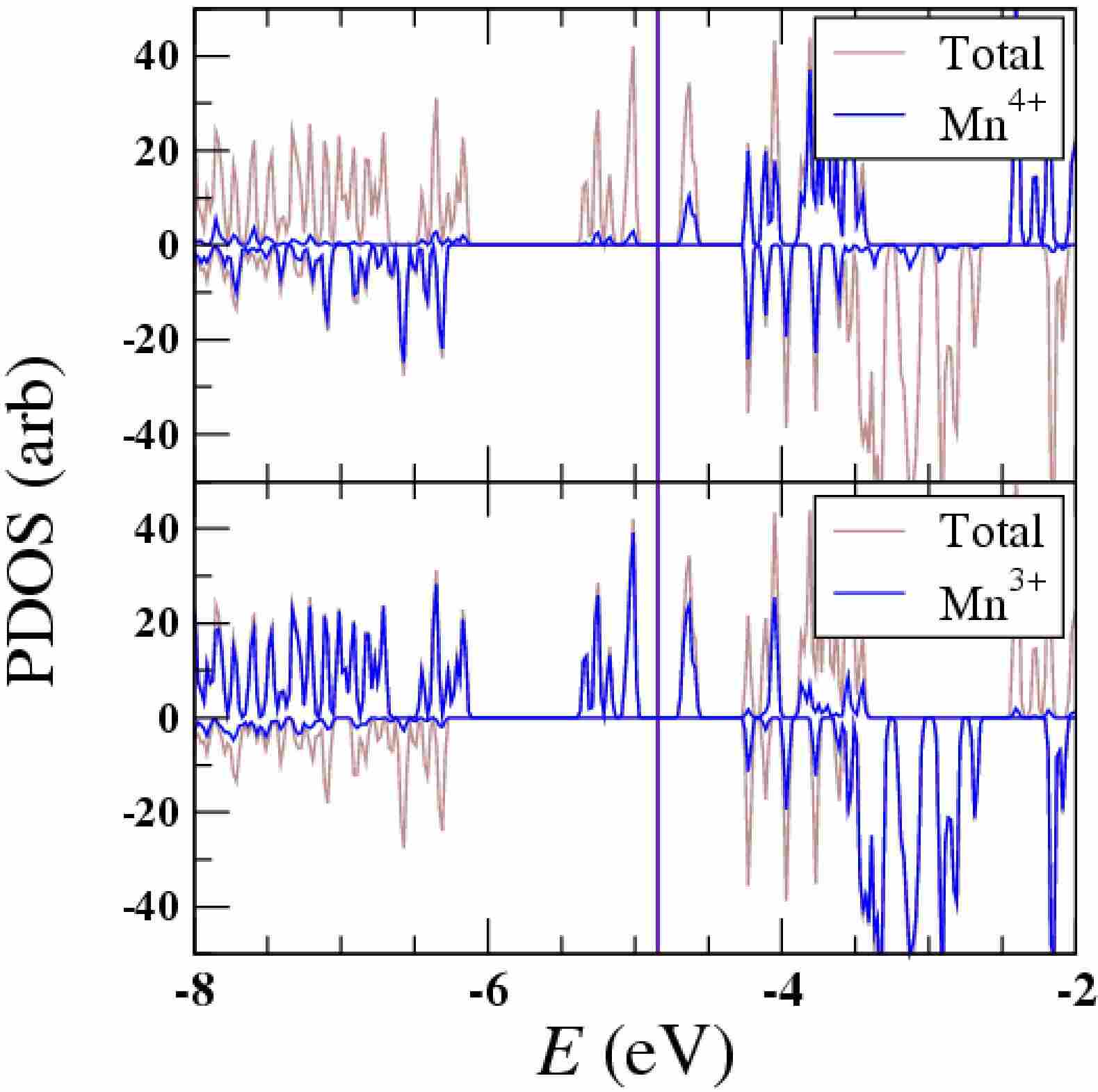}}
\caption{Orbital resolved density of states around the Fermi level (purple vertical line) for 
[Mn$_{12}$O$_{12}$(CH$_3$COO)$_{16}$(H$_2$O)$_4$]  in the $S=10$ ground state.
Note that the HOMO-LUMO gap is an intra-$d$ gap with states originating mainly from
Mn$^{3+}$. }
\label{Fig2}
\end{figure}
%\newpage
\section{Mn$_{12}$ Density of States}

In figure \ref{Fig1} we present the orbital resolved DOS for [Mn$_{12}$O$_{12}$(CH$_3$COO)$_{16}$(H$_2$O)$_4$]
as calculated with DFT-GGA for the $S=10$ ground state. We note that the HOMO-LUMO gap is an intra-$d$ gap formed
mainly between molecular orbitals with amplitude over the Mn ions. These have a large degree of hybridization with O 
and they are fully polarized around the Fermi level ($E_\mathrm{F}$). Also note that there is little contribution to the DOS 
around $E_\mathrm{F}$ from the C $p$ orbitals. 

In figure~\ref{Fig2} we further analyze the DOS by projecting it over the Mn$^{3+}$ and Mn$^{4+}$ ions individually. 
The figure demonstrate that the HOMO-LUMO gap is indeed solely determined by Mn$^{3+}$ $d$-shell with the first
of the Mn$^{4+}$ states being at least 1~eV away from $E_\mathrm{F}$.

\section{Transport simulation cell}

\begin{figure}[ht]
\centerline{{\epsfxsize=0.5\textwidth \epsffile{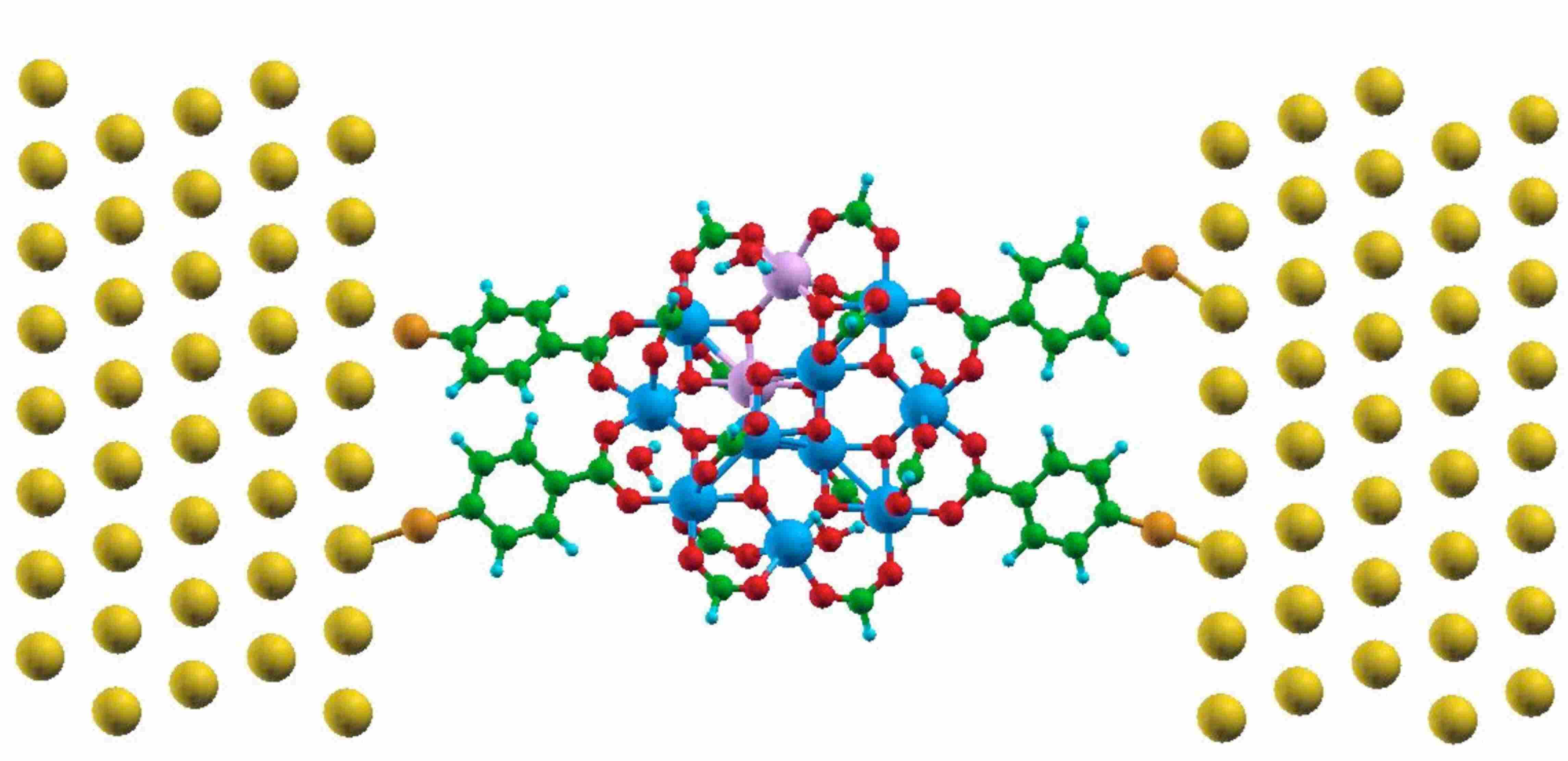}}}
\caption{Transport simulation cell used in this work. Color code: Blue=Mn, Red=O, Green=C, Light Blue=H, Yellow=Au,
Dark Yellow=S, Purple=Mn (flipped).}
\label{Fig3}
\end{figure}

The transport simulation cell is constructed from a [Mn$_{12}$O$_{12}$(CH$_3$COO)$_{16}$(H$_2$O)$_4$] molecule
comprising 16 thiol-terminated C$_6$H$_4$ ligands. We cut and passivate 12 of those 16 ligands in order to reduce
the lateral size of the cell. Such final molecule is relaxed in vacuum by standard conjugate gradient method and then 
positioned at a minimal energy location on the Au (111) surface. The search for the minimum is conducted as follows. 
First we place the S atom of one of the remaining 4 ligands at the hollow site of the Au (111) surface [this is one of the
preferential bonding sides for thiol on Au (111) \cite{Cormac}], and then we rotate the molecule about this position in the search of the total 
energy minimum. The final device simulation cell (see figure Fig.~\ref{Fig2}) also includes five Au atomic planes on each
side of the molecule. The lateral dimensions of these are those of a $6\times4$ supercell constructed from
the primitive cell along the $fcc$ (111) direction. The unit cell for the electrodes is also a $6\times4$ supercell.
The resulting scattering region (where the self-consistent calculation takes place) is presented
in figure~\ref{Fig2}. Note that two of the 12 Mn atoms are colored in purple (the other ones are blue). These are the ions
whose spins are flipped in forming the spin-flip configuration of the molecule.

%
% ========================================================

\end{document}